\documentclass[12pt]{iopart}

\usepackage{amssymb}
\usepackage{graphicx}
\usepackage{multirow}
\usepackage{booktabs,siunitx}
\usepackage{booktabs}

\begin{document}

\title[]{Robust electronic and tunable magnetic states in Sm$ _{2} $NiMnO$ _{6} $ ferromagnetic insulator}

\author{
S. Majumder$^{a}$,
\
M. Tripathi$^{a}$,
\
I. P\'{i}\v{s}$^{b,c}$,
\
S. Nappini$^{c}$,
\
P. Rajput$^{d}$,
\
S. N. Jha$^{d}$,
\
R. J. Choudhary$^{a}$,
and
D. M. Phase$^{a}$
\\
$^{a}$UGC DAE Consortium for Scientific Research, Indore 452001, India\\
$^{b}$Elettra Sicrotrone Trieste S.C.p.A., S.S. 14-km 163.5, 34149 Basovizza, Trieste, Italy\\
$^{c}$IOM CNR, Laboratorio TASC, S.S. 14-km 163.5, 34149 Basovizza, Trieste, Italy\\
$^{d}$Beamline Development and Application Section, Bhabha Atomic Research Centre, Mumbai 400085, India\\}
\ead{$^{*}$ram@csr.res.in}

\begin{abstract}
Ferromagnetic insulators (FM-Is) are the materials of interest for new generation quantum electronic applications. Here, we have investigated the physical observables depicting FM-I ground states in epitaxial Sm$ _{2} $NiMnO$ _{6} $ (SNMO) double perovskite thin films fabricated under different conditions to realize different level of Ni/Mn anti-site disorders (ASDs). The presence of ASDs immensely influence the characteristic magnetic and anisotropy behaviors in SNMO system by introducing short scale antiferromagnetic interactions in predominant long range FM ordered host matrix. Charge disproportion between cation sites in form of $ Ni^{2+}+Mn^{4+} \longrightarrow Ni^{3+}+Mn^{3+} $, causes mixed valency in both Ni and Mn species, which is found insensitive to ASD concentrations. Temperature dependent photo emission, photo absorption measurements duly combined with cluster model configuration interaction simulations, suggest that the eigenstates of Ni and Mn cations can be satisfactorily described as a linear combination of the unscreened $ d^{n} $ and screened $ d^{n+1} \underline{L} $ ($ \underline{L} $: O 2\textit{p} hole) states. The electronic structure across the Fermi level (E$ _{F} $) exhibits closely spaced Ni $ 3d $, Mn $ 3d $ and O $ 2p $ states. From occupied and unoccupied bands, estimated values of the Coulomb repulsion energy ($ U $) and ligand to metal charge transfer energy ($ \Delta $), indicate charge transfer insulating nature, where remarkable modification in Ni/Mn $ 3d $ - O $ 2p $ hybridization takes place across the FM transition temperature. Existence of ASD broadens the Ni, Mn $ 3d $ spectral features, whereas spectral positions are found to be unaltered. Hereby, present work demonstrates SNMO thin film as a FM-I system, where FM state can be tuned by manipulating ASD in the crystal structure, while I state remains intact.
\end{abstract}

\section{INTRODUCTION}
Ferromagnetic-Insulators (FM-Is) are opening new avenues for next generation spintronic devices and spin wave information processing applications \cite{YKWakabayashi2019, CSohn2019, SJNoh2019, NSRogadol2005, JSBenitez2011, MPSingh2011}. The exchange field proximity effect from FM state and negligible effective charge current due to I state, make FM-I useful as spin filter \cite{JLumetzberger2020}, quantum topological system \cite{JDSau2010}, and dissipationless medium \cite{JFrantti2019}. Achieving control over the ground state in FM-Is can further extend the feasibility of these systems in aforementioned technological perspectives \cite{SJNoh2019}. 

In contemporary FM system, the magnetic ordering generally involves RKKY or double exchange interactions, which are associated with metallic (M) conductivity \cite{CSohn2019}. On the other hand, insulating state in a material generally coupled with antiferromagnetic (AFM) super exchange interactions \cite{CSohn2019}. This is why, access to aforementioned scenario in practical system is scarcely found. In this regard, double perovskites A$_2$B'B"O$_6$ (A: alkaline earth or rare earth cations, B', B": transition metal cations octahedrally coordinated by six oxygen anions) with Ni, Mn as B-site ions are exceptional for possessing sparse FM-I state \cite{NSRogadol2005, JSBenitez2011, MPSingh2011}. Moreover, recent observation of spin pumping behavior in A$_2$NiMnO$_6$ (ANMO) thin films \cite{YShiomi2014, KMallick2019}, has renewed research interest on Ni-Mn double perovskite FM-Is, for technological feasibility as well as for fundamental importance. Despite several research endeavors, the reliability and reproducibility of many physical aspects in ANMO family are still questionable which create hindrance towards its usage in device fabrication. It is predicted that the virtual hopping of electrons between half filled e$ _{g} $ and empty e$ _{g} $ orbitals of Ni and Mn ions respectively, through Ni-O-Mn $ \sim $180$ ^{0} $ super exchange pathways, leads to FM-I ground state in B-site cation ordered ANMO system \cite{GoodenoughKanamoril195559}. In early reports, contradicting explanations are found regarding the charge state of transition metal ions participating in magnetic interaction. For instance, Goodenough \textit{et al.}\cite{JBGoodenoughl1961} explained the observed FM behavior considering Ni$ ^{3+} $-O-Mn$ ^{3+} $ superexchange interactions. Whereas, Blasse \textit{et al.}\cite{GBlassel1965} claimed that FM ordering is due to Ni$ ^{2+} $-O-Mn$ ^{4+} $ superexchange interaction. On the other hand, discrepancies in magnetic and electronic properties are observed in ANMO synthesized under different conditions, which arise possibly because of distinct growth kinetics driven varying cation arrangement in the system \cite{WZYang2012, FGheorghiu2015}. Comparable ionic radii of Ni and Mn ions in ANMO causes the formation of anti-site disorder (ASD) practically inescapable. ASD in double perovskite takes the decisive role in stabilizing the magnetic and electronic ground states \cite{MGHernandez2001, DDSarma2001}. Nevertheless, upto what extent ASD can transform the functional properties in any general double perovskite material is not predictable due to lack of proper understanding. This is because, the fine tuning, precise quantification and distribution characterization of ASD in complex lattice environment like double perovskite system, is not a straightforward exercise.

A comprehensive analysis of electronic structure can disclose the valence state of ions liable for magnetic exchange interaction in ANMO system. Moreover, the evolution of electronic energy landscape across the magnetic transition holds the key to interpret the unique magnetic phase diagram. On the other hand, to figure out the ASD driven alteration in physical observable corresponding to electronic and magnetic ground states, a systematic study is needed on ANMO system having different cation arrangements. In this backdrop, we aim to investigate temperature dependent electronic and magnetic properties of epitaxial Sm$_{2}$NiMnO$_{6}$ (SNMO) thin films grown under different conditions to have different level of cation arrangements. Our findings suggest that SNMO double perovskite thin film represents a FM-I system, in which FM behavior can be controlled by proper tuning of ASD in the system, whereas the I nature is comparatively rigid. 

\section{EXPERIMENTAL METHODOLOGY}
Following the recipe reported in Ref.\cite{SMajumder2022}, epitaxial SNMO thin films (thickness $\sim$140$\pm$5 nm) were deposited on (001) oriented SrTiO$ _{3} $ single crystal substrates by pulsed laser deposition using a KrF excimer laser system (Lambda Physik, wavelength $ \lambda $=248 nm, pulse width 20 ns). For magnetization measurements MPMS 7-Tesla SQUID-VSM (Quantum Design Inc., USA) system was used. The average sensitivity in magnetic moment measurement was of the order of $ \sim 10^{-8}$ emu. Magnetization as a function of temperature M(T) and as a function of magnetic field M(H) were recorded by applying the measuring magnetic field $ \mu_{0} $H along in-plane direction of the thin films. Prior to measurements, trapped magnetic field inside the magnetometer superconducting magnet was nullified following standard de-Gaussing protocol and any magnetic history of the sample was erased by heating the films above their corresponding magnetic ordering temperatures. M(T) measurements were carried out in typical field cooled warming (FCW) cycle. The background (substrate and sample holder) diamagnetic contribution from M(H) curves was removed by standard high field linear slope subtraction method. Core level X-ray photo electron spectroscopy (XPS) studies were carried out at T=300 K temperature, using Omicron (EA-125, Germany) hemispherical energy analyzer and Al $ K_{\alpha} $ ($ h\nu $=1486.7 eV) X-ray source at Angle Integrated Photoemission Spectroscopy (AIPES) beamline (Indus-1, BL 2, RRCAT, Indore, India). Ultra high vacuum (UHV) having base pressure of the order of $ \sim $10$ ^{-10} $ Torr, was maintained in the experimental chamber during the XPS measurements. The charging effect corrections in observed binding energies were done by recording XPS of C 1\textit{s} specie (present on the surface of the films). The estimated energy resolution for the photon energy range used in XPS studies was found to be $ \sim $0.6 eV. To explore the electronic interactions, configuration interaction cluster model charge transfer multiplet (CTM) simulations were performed employing CTM4XAS program \cite{HIkeno2009}. Resonant photo emission spectroscopy (ResPES) and X-ray absorption near edge spectroscopy (XANES) experiments were carried out at BACH beamline (8.2, Elettra synchrotron facility, Trieste, Italy) \cite{MZangrando2001, MZangrando2004}. ResPES experiments were conducted at T=300 K, 125 K and 100 K temperatures, by measuring valence band photo emission spectra while sweeping the incident photon energy through Ni, Mn $ 2p_{3/2} \rightarrow 3d $ photo excitation thresholds. Photo emission spectra were recorded in normal emission geometry using a VG-SCIENTA R3000 hemispherical analyzer inclined at 60$ ^{o} $ from the incident beam direction. In photo emission measurements, the binding energies were calibrated to the main C 1\textit{s} photo emission line ($ \sim $284.8 eV) of adventitious carbon and the energy resolution was $ \sim $0.3 eV. Analysis and deconvolution of observed spectra were done using Igor Pro and XPSPEAK 4.1 programs. XANES measurements were carried out at T=300 K, 125 K and 100 K temperatures, across the O $ K $ photo absorption edge. Photo absorption spectra were acquired in total electron yield detection mode by measuring the drain current using 428 Keithley electrometer. The energy calibration in photo absorption spectra was performed by measuring Au $ 4f $ photo emission spectra on a gold reference sample. The estimated energy resolutions for these experiments were found to be $ \sim $ 0.08 eV, 0.1 eV and 0.2 eV for the spectral characters corresponding to energy of $ \sim $530 eV, 640 eV and 853 eV, respectively. Observed XANES spectra were corrected following standard normalization procedures using ATHENA program \cite{BRavel2005}. For aforementioned spectroscopic measurements, the incident beam polarization was fixed to linear horizontal. The multi-spectroscopy experimental chamber was at UHV condition with a base pressure of the order of $ \sim10^{-10} $ Torr. In order to get better statistics and confirm the data reproducibility, all the spectroscopic scans were collected several times and merged. Temperature dependent Ni $ K $ XANES were recorded in fluorescence mode using Vortex detector and hard X-ray synchrotron radiation at Scanning EXAFS beamline (BL 9, Indus-2, RRCAT, Indore, India). Standard normalization procedures were applied on XANES spectra using ATHENA program. Incident X-ray beam energy calibration was conducted by measuring reference spectra of Ni metal foil. Energy resolution for Ni $ K $ edge XANES measurements was found to be $ \sim $0.8 eV.

\section{RESULTS AND DISCUSSION}
The growth conditions during the fabrication of SNMO thin films, as discussed in Ref.\cite{SMajumder2022}, are exploited as the tuning knobs to effectively tailor the Ni/Mn cation arrangements in the system. Real time \textit{in situ} reflection high energy electron diffraction and high resolution X-ray diffraction rocking $ \omega $ scans suggest similar growth mode, similar surface morphology, and similar dislocation defect densities in different SNMO films irrespective of having different growth conditions. Structural characterization utilizing X-ray diffraction 2$ \theta $ scans, and reciprocal space mapping (RSM) confirm the single phase (SG: \textit{P2$ _{1} $/n}) epitaxial growth of SNMO thin films oriented along (001) direction. RSM result confirms that the grown films are relaxed from substrate clamping effects and consequently, the lattice parameters of SNMO crystals in the thin films along in-plane and out-of-plane directions are found close to it's bulk counterparts. The degree of ASD concentration and their distribution morphology are probed at both microscopic and macroscopic level implementing extended X-ray absorption fine structure, bulk magnetometry and random alloy structure simulations. The details about aforementioned structural characterizations are discussed elsewhere \cite{SMajumder2022}. In present work, to explore the role of varying ASD on magnetic and electronic states of SNMO system, two samples are selected as the representing platform, (i) more ordered thin film, namely S$ \_ $HO with ASD density Q$_{ASD}\sim5\pm1\%$ and (ii) more disordered thin film namely S$ \_ $I with Q$_{ASD}\sim19\pm2\%$. It is important to mention here that neither fully ordered (Q$_{ASD}=0\%$) nor fully disordered (Q$_{ASD}=50\%$) situation can be achieved experimentally in real crystal system. This is due to very small difference in ionic radii of Ni and Mn ions and hence, presence of ASD is intrinsic in SNMO system. From our previous study \cite{SMajumder2022}, we have observed that for low ASD concentration, disordered structures are found to be homogeneously distributed in the ordered host matrix. 

\begin{figure*}[t]
\centering
\includegraphics[angle=0,width=0.8\textwidth]{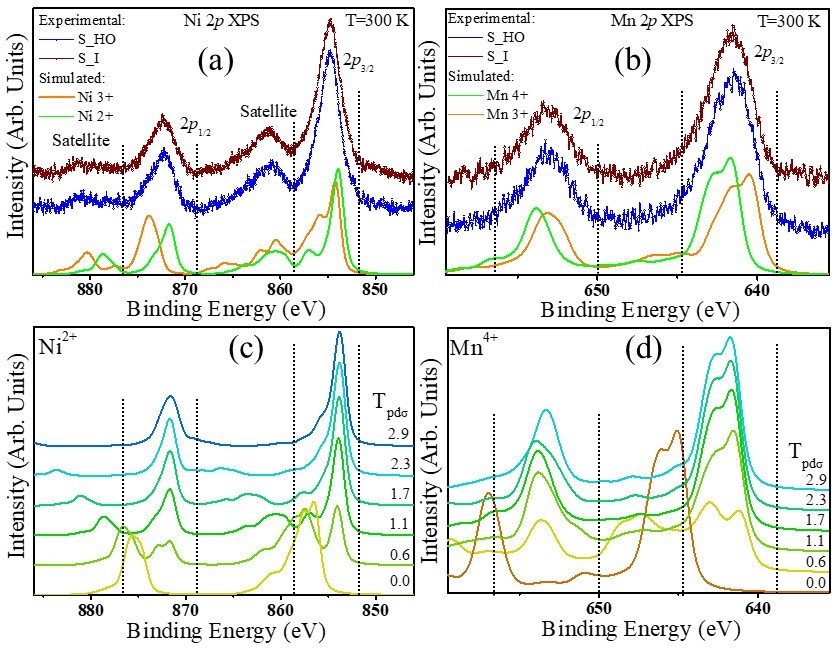}
\caption{Experimentally observed and simulated $ 2p $ core level photo emission spectra measured at T=300 K for (a): Ni and (b): Mn cations present in SNMO thin films having different anti-site disorder densities. Simulated spectra for (c): Ni$ ^{2+} $ and (d): Mn$ ^{4+} $ species at T=300 K, with varying metal - ligand hybridization strength. The vertical dotted lines are for guide to the eyes to track the positions of spectral features. To have better visualization spectra are vertically translated here.}\label{xps}
\end{figure*}

\begin{figure*}[h!]
\centering
\includegraphics[angle=0,width=0.8\textwidth]{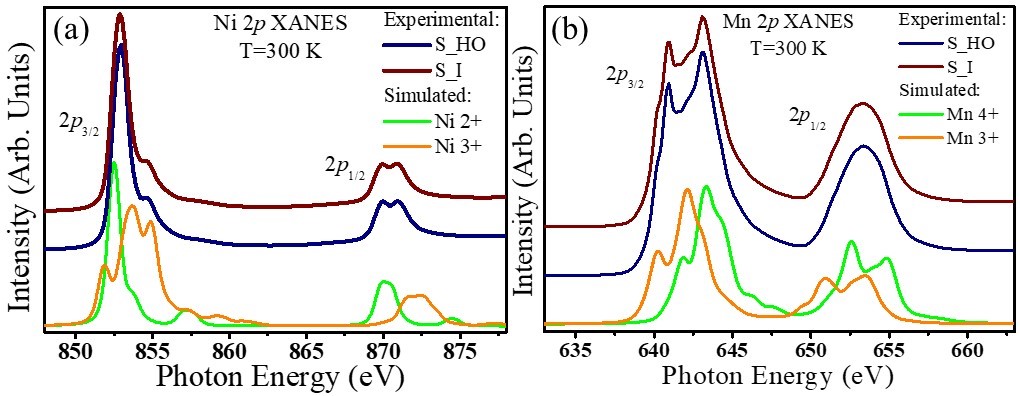}
\caption{Experimentally observed and simulated 2\textit{p} core level photo absorption spectra measured at T=300 K for (a): Ni and (b): Mn cations present in SNMO thin films having different anti-site disorder densities. To have better visualization spectra are vertically translated here.}\label{xas300k}
\end{figure*}

Ni and Mn $ 2p $ core level XPS spectra measured at T=300 K for the SNMO thin films display spin-orbit splitted low lying $ 2p_{3/2} $ and higher lying $ 2p_{1/2} $ states along with the corresponding satellite structures, as presented in Figs. \ref{xps}(a, b). Inelastic background contributions are removed from recorded spectra by subtracting Shirley function. Ni and Mn 2\textit{p} $\rightarrow$ 3\textit{d} (\textit{L}) edge XANES spectra splitted in low lying $ 2p_{3/2} $ (\textit{L}$ _{3} $) and higher lying $ 2p_{1/2} $ (\textit{L}$ _{2} $) multiplets due to spin-orbit interaction, recorded at T=300 K for SNMO films, are displayed in Figs. \ref{xas300k}(a, b). In perovskite nikelets and manganites, temperature is observed to drive the valence state transition \cite{MNaka2016, NMannella2008}. It is necessary to check the possibility of such temperature induced valence transition in SNMO double perovskite. Ni and Mn \textit{L} edge XANES spectra acquired for sample S$\_$HO across the magnetic ordering temperature T$ _{C} $ are shown in Figs. S1(a, b) in Supplemental Material (SM). Here, the slight relative shift with respect to temperature variation, which is under the instrumental resolution limit, can be ignored. We have not observed any significant temperature driven changes in the spectral position or shape for both Ni and Mn species. Therefore, valence state across the temperature T$ _{C} $, remains same. We have also confirmed this from temperature dependent Ni \textit{K} edge XANES measurements (as shown in Fig. S7 in SM). 

To explore the nature of electronic interactions, configuration interaction cluster model charge transfer multiplet (CTM) simulations were performed on the core level $ 2p $ XPS and $ 2p $ XANES spectra employing CTM4XAS program \cite{HIkeno2009}. Accounting the cluster model approximation, the ground state wave function for the late transition metal central cation with formal $ d^{n} $ configuration surrounded by ligand anions can be considered as \cite{AEBocquet1992I}, 
\begin{equation}\label{EqCMg}
\Psi_{g} = a_{0} \vert \textit{d}^{n} \rangle + \sum_{m} a_{m} \vert \textit{d}^{n+m} \underline{L}^{m} \rangle : m = 1,..10-n
\end{equation}
where $ \underline{L} $ denotes a hole in the ligand band. Such mixing of pure ionic ($ d^{n} $) and charge transfer screened ($ d^{n+m} \underline{L}^{m} $) characters at the ground state is experimentally confirmed in several transition metal oxide compounds \cite{LHTjeng1991, TSaitoh1995II, OTjernberg1996I}. Applying the Slater approximation, the Coulomb and charge transfer interactions are described by the diagonal Hamiltonian matrix elements of $ d^{n} $ ion basis states \cite{AEBocquet1992I}, 
\begin{eqnarray}\label{EqdiaH}
 & \langle d^{n+m} \underline{L}^{m} \vert H \vert d^{n+m} \underline{L}^{m} \rangle = \nonumber\\
 & \varepsilon (d^{n+m} \underline{L}^{m}) - \varepsilon (d^{n}) + m \Delta + \frac{1}{2} m(m-1)U,
\end{eqnarray}
where $ \Delta $ represents the ligand to metal charge transfer energy, 
\begin{equation}\label{EqDpd}
\Delta_{p-d} = E(d^{n+1} \underline{L}) - E(d^{n}),
\end{equation}
and $ U $ corresponds to the Coulomb repulsion energy, 
\begin{equation}\label{EqUdd}
U_{d-d} = E(d^{n-1}) + E(d^{n+1}) - 2E(d^{n}).
\end{equation}
Transition metal $ d $ - ligand $ p $ hybridization is related to off diagonal matrix elements as \cite{AEBocquet1992I},
\begin{equation}\label{EqT}
T = \langle d_{\alpha} \vert H \vert L_{\alpha} \rangle,
\end{equation}
where $ d_{\alpha} $ and $ L_{\alpha} $ symbolize electron from transition metal and ligand bands respectively, with same orbital symmetry. Octahedral ligand field splits the transition metal $ d $ bands into $ t_{2g} $ and $ e_{g} $ states with energy separation defined as the crystal field splitting energy $10Dq$ \cite{AEBocquet1992I}. $ T $ is anisotropic that is $ T_{e_{g}} \neq T_{t_{2g}} $. In the octahedral geometry hybridization strengths are expressed in terms of the Slater–Koster transfer integrals, as $ T_{e_{g}} = \sqrt{3} T_{pd\sigma} $, $ T_{t_{2g}} = 2 T_{pd\pi} $, where $ T_{pd\sigma}/T_{pd\pi} \approx -2.2 $ as reported in previous studies \cite{AEBocquet1992I}. For $ 2p $ core level photo emission and photo absorption process, a core hole is created in $ 2p $ level, which influences the final state and causes a constant reduction in final state energy. This interaction is described as $ 3d $ electron - $ 2p $ core hole attractive potential $ U_{dc} $. In case of $ 2p $ core level XPS, the final state is described as \cite{AEBocquet1992I},
\begin{equation}\label{Eq2ppee}
\Psi_{f} = \widetilde{a}_{0} \vert \underline{c} \textit{d}^{n} \rangle + \sum_{m} \widetilde{a}_{m} \vert \underline{c} \textit{d}^{n+m} \underline{L}^{m} \rangle : m = 1,..10-n,
\end{equation}
where $ \underline{c} $ refers to a core hole. For $ 2p \rightarrow 3d $ XANES, the final state is represented by, 
\begin{equation}\label{Eq2ppae}
\Psi_{f} = \acute{a}_{0} \vert \underline{c} \textit{d}^{n+1} \rangle + \sum_{m} \acute{a}_{m} \vert \underline{c} \textit{d}^{n+m+1} \underline{L}^{m} \rangle : m = 1,..10-n.
\end{equation}
The core level $ 2p $ XPS spectra is calculated by applying sudden approximation as \cite{AEBocquet1992I},
\begin{equation}\label{Eq2ppes}
\rho(e_{k}) = \sum_{f} \vert \langle \Psi_{f} \vert c \vert \Psi_{g} \rangle \vert^{2} \delta(h\nu-e_{k}-E_{f}),
\end{equation}
where $ c, h\nu, e_{k} $ and $ E_{f} $ represent annihilation process of core electron, photon energy, photoelectron kinetic energy and final state energy, respectively. Similarly, using sudden approximation the $ 2p $ XANES spectra is computed by \cite{FMFdeGroot1990II},
\begin{equation}\label{Eq2ppas}
\sigma(h\nu) = \sum_{f} \vert \langle \Psi_{f} \vert p \vert \Psi_{g} \rangle \vert^{2} \delta(h\nu+E_{i}-E_{f}),
\end{equation}
where $ p $ and $ E_{i} $ denote photon absorption process and initial state energy respectively. Corresponding intensities of core level XPS and XANES spectra are found proportional to $ \vert a_{0} \widetilde{a}_{0} + \sum_{m} a_{m} \widetilde{a}_{m} \vert^{2} $ and $ \vert a_{0} \acute{a}_{0} + \sum_{m} a_{m} \acute{a}_{m} \vert^{2} $, respectively \cite{TMizokawa1995I}. 

To simplify the computations, we have ignored higher order charge transfer states ($ d^{n+m} \underline{L}^{m} $: m$ \geqslant $2, where $ \underline{L} $ denotes a hole in ligand band), as they usually have low contributions. Parameters used in CTM simulations for SNMO system are listed in Table \ref{tabCTM}. Here, $ \Delta_{p-d} $, $ U_{d-d} $, $ U_{dc} $, $ 10Dq $, and $ T_{pd\sigma} $ represent: ligand to metal charge transfer energy, Coulomb repulsion energy, $ 3d $ electron - $ 2p $ core hole attractive potential, crystal field splitting energy, and transition metal $ d $ - ligand $ p $ hybridization strength, respectively. The values of these parameters are obtained from experimental electronic band structure analysis (discussed later). To consider configuration interaction effects, the Slater integrals ($ 3d3d $, $ 2p3d $) are suppressed to 80$ \% $ of the Hartree-Fock values \cite{AEBocquet1992I}. To take into account the broadening in observed spectra, a convolution between Lorentzian and Gaussian line shapes with full width at half maximum (FWHM) 0.6 eV and 0.8 eV respectively, is considered. However, various intrinsic as well as extrinsic complex mechanisms involved in measurement process result in greater broadening of experimental profile than simulated spectra \cite{FMFdeGroot1990II}. 

CTM simulated $ 2p $ XPS and XANES spectra for Ni$^{2+/3+}$ and Mn$^{4+/3+}$ species are presented in Figs. \ref{xps}(a, b) and Figs. \ref{xas300k}(a, b). Comparison of experimental and simulated spectra suggests mixed valency of both Ni, Mn ions. Sm 3\textit{d} XPS spectra for the SNMO thin film system are depicted in Fig. S2 in SM. The observed Sm 3\textit{d} photo emission spectral shapes match very well with Sm$ _{2} $O$ _{3} $ (Sm valency: 3+) XPS data \cite{AReisner2017}. Comparing the 3\textit{d} peak binding energy positions (3\textit{d}$ _{5/2} \sim $1084.1 eV, 3\textit{d}$ _{3/2} \sim $1111.2 eV), it is confirmed that Sm in SNMO is in 3+ valence state only. We have ruled out the possible presence any major oxygen vacancy defects in SNMO thin films (discussed later). Deconvolution analysis of spectroscopic results as discussed in Ref. \cite{SMajumder2022}, confirm that in both S$ \_ $HO and S$ \_ $I samples, Ni and Mn transition metals have mixed valence nature with fractional concentrations: 55$ \pm 1 \% $ Ni$ ^{2+} $, 45$ \pm 1 \% $ Ni$ ^{3+} $ and 54$ \pm 2 \% $ Mn$ ^{4+} $, 46$ \pm 2 \% $ Mn$ ^{3+} $. In early reports on bulk La$ _{2} $NiMnO$ _{6} $ (LNMO) system, contradicting explanations were found whether the charge state of transition metal ions are Ni$ ^{3+} $, Mn$ ^{3+} $ \cite{JBGoodenoughl1961, CLBull2003}; or Ni$ ^{2+} $, Mn$ ^{4+} $ \cite{GBlassel1965, JBlasco2002}. On the other hand, in LNMO thin film it was observed that along with dominating 4+ valency, Mn species can also have small amount of 3+ charge state contributions \cite{MKitamura2009}. From our studies we have observed that in SNMO system, Ni and Mn transition metals have mixed valence nature. The observed mixed valency is originated due to charge disproportionation between transition metal sites through $ Ni^{2+}+Mn^{4+} \longrightarrow Ni^{3+}+Mn^{3+} $. Similar kind of charge disproportionate behavior is also observed in several other double perovskites \cite{GHJonker1966, JBGoodenough2000}. The position of core level features remains unaffected with respect to variation in ASD concentrations, which suggests same chemical valence state of constituting elements present in both of the SNMO thin films.  

\begin{table}[]
\centering
\begin{tabular}{@{}c|c|c@{}}
\hline
Parameters & Ni (eV) & Mn (eV) \\ 
\hline
$ \Delta_{p-d} $        & 2.4 & 3.5 \\
$ U_{d-d} $        & 5.6 & 5.0 \\
$ U_{dc} $        & 7.6 & 7.0 \\
$ 10Dq $       & 1.0 & 1.5 \\
$ T_{pd\sigma} $       & 1.1 & 1.7 \\
\hline
\hline
\end{tabular}
\caption{Parameters used in Ni, Mn $ 2p $ core XPS and XAS spectra simulations for SNMO system.}
\label{tabCTM}
\end{table}

In transition metal compounds with moderate ionic and covalent nature, the main peak and satellite feature of $ 2p $ XPS are generally attributed to low energy $ \vert \underline{c} d^{n+1} \underline{L} \rangle $ (where $ \underline{c} $ refers to a core hole), and higher energy $ \vert \underline{c} d^{n} \rangle $ characters, respectively \cite{AEBocquet1992I}. In CTM simulation, Ni/Mn $ 3d $ - O $ 2p $ hybridization strength is optimized to adequately reproduce the spectral position and weight of satellite feature with respect to main peak, as observed in experimental spectra. The evolution of $ 2p $ core XPS spectra with varying hybridization parameter T$ _{pd\sigma} $ are illustrated in Figs. \ref{xps}(c, d). With increasing metal - ligand hybridization strength from T$ _{pd\sigma} $=0, the satellite structure corresponding to main peak emerges at finite T$ _{pd\sigma} $ value. The appearance of satellite features in CTM simulations in accordance with experimental profile confirms the reliability of charge transfer approach with proper simulation parameters. Therefore, CTM analysis on SNMO system reveals that there is a significant hybridization between transition metal $ d $ and ligand $ p $ orbitals where unscreened ($ d^{n} $) and screened ($ d^{n+1} \underline{L}^{1} $) states are mixed together. Similar mixing of unscreened and screened characters at the ground state is experimentally confirmed in several transition metal oxide compounds \cite{LHTjeng1991, TSaitoh1995II}.

\begin{figure*}[t]
\centering
\includegraphics[angle=0,width=1.0\textwidth]{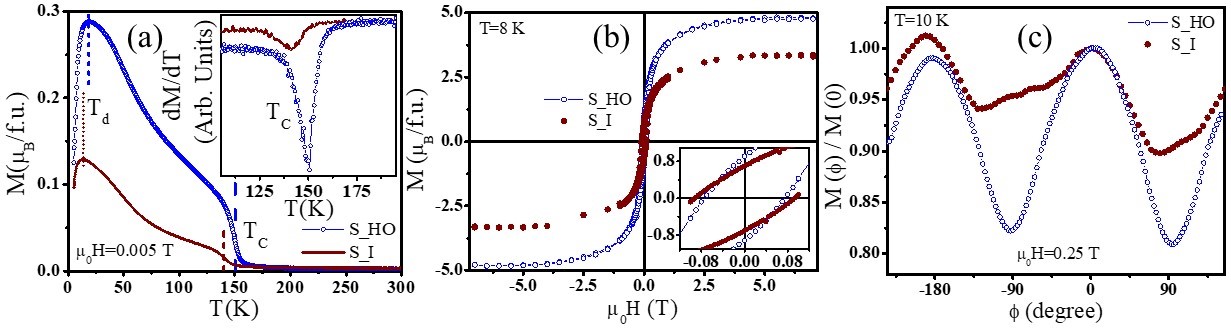}
\caption{Magnetometric measurements on SNMO thin films with different anti-site disorder concentrations. (a): Temperature dependent magnetization M(T) recorded under field cooled warming mode with applied magnetic field $ \mu_{0} $H=0.005 T along in-plane geometry. Inset: First order temperature derivative of magnetization dM/dT. (b): Isothermal magnetization M(H) measured as a function of applied field along in-plane geometry at T=8 K. Inset: Enlarged view of low field region. (c): Angular dependency of magnetization M($ \phi $) acquired at T=10 K under $ \mu_{0} $H=0.25 T measuring magnetic field.}\label{mtmhmth}
\end{figure*}

Magnetization measured as a function of temperature M(T) under FCW cycle with applied magnetic field of $ \mu_{0} $H=0.005 T for the SNMO thin films are presented in Fig. \ref{mtmhmth}(a). Observed magnetic transitions are nomenclatured as follows: (i) the onset in M(T) at T$ _{C} $=150.1 K for S$\_$HO, T$ _{C} $=140.1 K for S$\_$I; (ii) the downturn in M(T) at T$ _{d} $=17.7 K for S$\_$HO, T$ _{d} $=14.2 K for S$\_$I; and (iii) the broad inverted cusp like behavior within T$_{d}<$T $<$T$_{C}$. The interplay between various possible magnetic exchange pathways causes different temperature driven magnetic orderings in SNMO system. With decreasing temperature, observed first transition at T=T$ _{C} $ is ascribed as paramagnetic (PM) to long range FM ordering of Ni-Mn cation ordered sublattices due to Ni-O-Mn superexchange interaction \cite{NSRogadol2005, SMajumder2022b}. The low temperature transition across T=T$ _{d} $ is possibly due to antiparallel coupling between polarized PM moments of Sm sublattice and FM moments of Ni-Mn network \cite{JSBenitez2011, SMajumder2022b}. The presence of ASD introduces short range AFM interactions through Ni-O-Ni, Mn-O-Mn superexchange pathways. The inverted cusp behavior in between T$_{d}<$T $<$T$_{C}$ temperature regime is attributed to signatures of ASD present in SNMO system \cite{MPSingh2011, SMajumder2022b}. Cation order and disorder structures mediated intercompeting long range FM and short range AFM interactions govern the magnetic properties in SNMO. With increasing ASD concentrations, broadening and shift in magnetic transitions toward low temperature values (Inset of Fig. \ref{mtmhmth}(a)) are observed. Isothermal magnetization M(H) for the SNMO samples acquired at T=8 K shows hysteresis behavior with respect to magnetic field cycling, as depicted in Fig. \ref{mtmhmth}(b). From measured M(H) curves, saturation moment (at $ \mu_{0} $H=70 kOe) and remanence values (Inset of Fig. \ref{mtmhmth}(a)) are found to be M$ _{S} $(S$ \_ $HO) $ \sim $4.8 $ \mu_{B} $, M$ _{S} $(S$ \_ $I) $ \sim $3.3 $ \mu_{B} $ and M$ _{R} $(S$ \_ $HO) $ \sim $0.9 $ \mu_{B} $, M$ _{R} $(S$ \_ $I) $ \sim $0.7 $ \mu_{B} $, respectively. With increasing ASD densities, drastic drop in saturation magnetization as well as remanence moment values are observed. Assuming the ideal Ni, Mn cation ordered case, the estimated $ M_{cal} $ values from different spin configurations of Sm$ ^{3+} $, Ni$ ^{2+/3+} $ and Mn$ ^{4+/3+} $ ions, yield $ \sim $5.12 $\mu_{B}$ and 5.62 $\mu_{B}$ for Ni$^{3+}$ low spin and Ni$^{3+}$ high spin states, respectively \cite{SMajumder2022}. The reduction in the observed moment value in SNMO systems can be understood by considering the presence of significant antiferromagnetic couplings related to anti-site cation disorders in background of ferromagnetic cation ordered host matrix. For cation ordered LNMO system, the reported value of saturation moment at 5 K temperature in presence of applied magnetic field of 5 T, was 4.96 $ \mu_{B} / f.u. $ \cite{NSRogadol2005}. Whereas, in case of highly cation ordered SNMO, sample S$ \_ $HO (with ASD density Q$ _{ASD} = 5 \pm 1 \% $) possess M$ _{S} $(at T = 5 K, $ \mu_{0} $H = 5 T) = 4.92 $ \mu_{B} / f.u. $ This small difference in saturation magnetization may arises because of change in A-site rare-earth ion. Angular dependency of magnetization M($ \phi $) measured at T=10 K, $ \mu_{0} $H=0.25 T depict dominating uniaxial characters of anisotropy with half of the periodicity values $ \sim $95.4$ ^{0} $ for S$ \_ $HO and 79.9$ ^{0} $ for S$ \_ $I, as presented in Fig. \ref{mtmhmth}(c). Rise in ASD fractions causes decrease in anisotropy and the M($ \phi $) curves become asymmetric leading to deviation from uniaxial nature to biaxial like anisotropy behavior. Therefore, the magnetic states in SNMO system can be effectively tuned by controlling Ni/Mn cation arrangement in the host matrix.

\begin{figure*}[t]
\centering
\includegraphics[angle=0,width=1.0\textwidth]{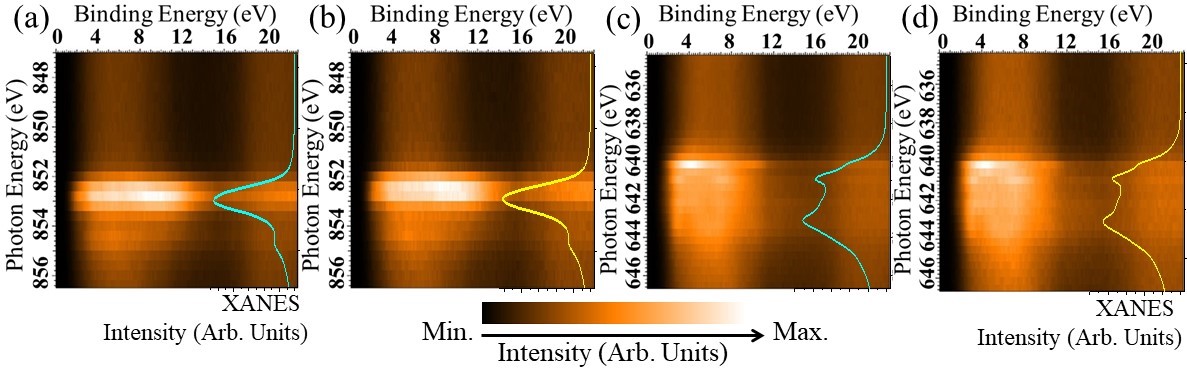}
\caption{Resonant photo emission maps along with near edge photo absorption spectra for (a): S$ \_ $HO thin film across Ni, (b): S$ \_ $I thin film across Ni, (c): S$ \_ $HO thin film across Mn and (d): S$ \_ $I thin film across Mn $ 2p_{3/2} \rightarrow 3d $ excitation thresholds acquired at T=300 K.}\label{respesmap}
\end{figure*}

To identify element specific spectral characters, present in valence band electronic structure, resonant photoemission technique is utilized. When the incident photon energy ($h\nu$) is tuned to excitation gap between transition metal $ lp \rightarrow 3d$ level ($ l $ = 2 or 3), photoemission from valence band is affected (enhanced or reduced). This effect can be understood considering two different photoemission process occurring simultaneously, (i) direct photoemission ($ 3d^{n} + h\nu \rightarrow 3d^{n-1} + e $) and (ii) photo absorption followed by super Coster-Kronig Auger decay ($ lp^{6} 3d^{n} + h\nu \rightarrow [lp^{5} 3d^{n+1}]^{*} \rightarrow lp^{6} 3d^{n-1} + e $, where the asterisk denotes intermediate excited state) \cite{GvanderLaan1992I, MWeinelt1997, LSangaletti2003}. For shallower $3p$ threshold these two photoemission channels have same order of magnitudes and interfere in complex fashion. Moreover, $ 3p $ spin orbit splitting is small in comparison to core hole lifetime and observed resonance enhancement is weak. On the other hand, in case of $2p$ threshold, these issues have lesser impact. At $2p$ regime, usually giant resonance is observed in which Coster-Kronig deexcitation channel overwhelms direct photoemission channel causing negligible interference effects. Consequently, observed resonance behavior clearly evidences $ d^{n-1} $ final state valence band features \cite{LHTjeng1991}. Therefore, $ 2p $ ResPES study is a more straight forward methodology than $ 3p $ ResPES, to conclusively determine the contribution of $ 3d $ character in valence band. Furthermore, as 2\textit{p} ResPES measurement involves higher energy photon excitation than 3\textit{p} case, provides information about more penetration depth from the sample surface.

\begin{figure*}[t]
\centering
\includegraphics[angle=0,width=0.8\textwidth]{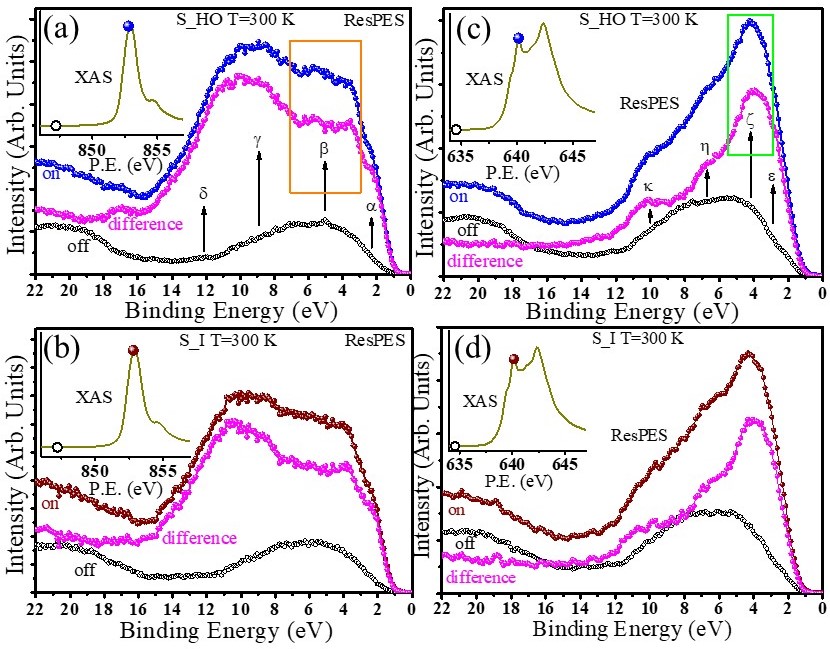}
\caption{On, off resonance and on-off difference valence band photo emission spectra for (a): S$ \_ $HO sample at Ni, (b): S$ \_ $I sample at Ni, (c): S$ \_ $HO sample at Mn and (d): S$ \_ $I sample at Mn $ 2p_{3/2} \rightarrow 3d $ absorption thresholds recorded at T=300 K. Insets: The on, off resonance photon energies are highlighted in corresponding $ 2p_{3/2} \rightarrow 3d $ photo absorption edges.}\label{onoffrespes}
\end{figure*}

For SNMO films, Figs. \ref{respesmap}(a-d) present ResPES mapping along with XANES spectra recorded at T=300 K, across Ni and Mn $2p_{3/2} \rightarrow 3d$ ($ L_{3} $) thresholds, respectively. Valence band characters show distinct resonance behaviors at corresponding absorption edge energies. In order to have better insight on intensity variation as well as energy dispersion of valence band features, spectrum for different incident photon energies are stacked vertically as shown in Figs. S3-S4 in SM. Absence of spectral weights at E$ _{F} $ (binding energy = 0 eV) suggests insulating nature of the system. Across Ni and Mn $ L_{3} $ absorption edge, maximum intensity enhancement in valence band spectra is observed at $h\nu$=852.8 eV and $h\nu$=640.1 eV, respectively. The on and off resonance spectra for Ni and Mn $ L_{3} $ ResPES along with corresponding photo absorption edges are presented in Figs. \ref{onoffrespes}(a-d). The photon energies for on, off resonances are highlighted in XANES spectra, as depicted in insets of Figs. \ref{onoffrespes}(a-d). Point by point on-off resonance difference spectra measures transition metal $ 3d $ characters present in valence band \cite{LHTjeng1991}. From observed resonance behavior, the valence band spectral lines are identified as follows: features at binding energy $ \sim $2.3 eV ($ \alpha $), 5.0 eV ($ \beta $), 8.9 eV ($ \gamma $), 12.2 eV ($ \delta $) are originated from Ni states; whereas $ \sim $2.9 eV ($ \varepsilon $), 4.2 eV ($ \zeta $), 6.8 eV ($ \eta $), 10.1 eV ($ \kappa $) binding energy features belong to Mn states. 

\begin{figure*}[t]
\centering
\includegraphics[angle=0,width=0.8\textwidth]{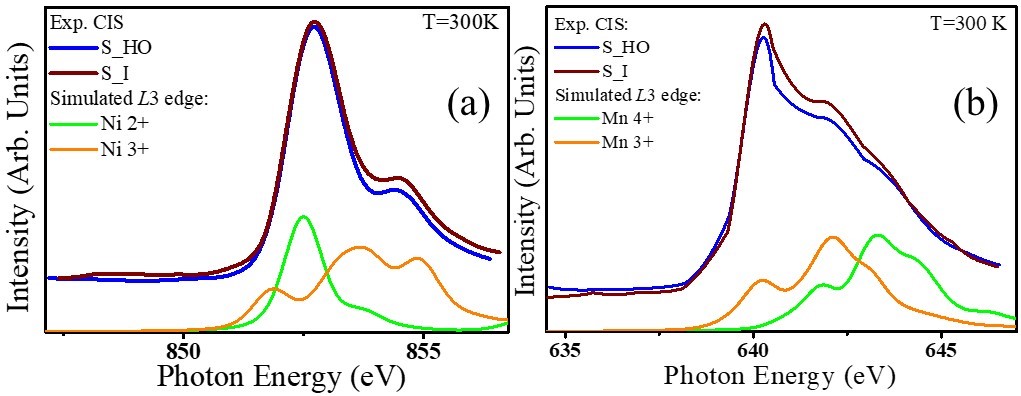}
\caption{Experimental constant initial state profile corresponding to (a): 5.0 eV, (b): 4.2 eV binding energy valence band characters recorded across Ni, Mn $ 2p_{3/2} \rightarrow 3d $ absorption thresholds respectively, for SNMO thin films having different degree of anti-site disorders along with simulated $ 2p_{3/2} \rightarrow 3d $ photo absorption spectra. To have better visualization spectra are vertically translated here.}\label{cisxasl3}
\end{figure*}

The spectral features which show large intensity increment across $ L_{3} $ thresholds are defined here as the main resonance lines. Ni and Mn main resonance features $ \beta $ and $ \zeta $ respectively, are highlighted by box in Figs. \ref{onoffrespes}(a, c). The feature $ \gamma $ observed in Ni $ L_{3} $ ResPES is not considered as main resonance line, as it also has Auger emission contributions (discussed later). In order to confirm the nature of main resonance lines, constant initial state (CIS) profile are studied. CIS plot is obtained from integrated intensity variation for a given binding energy feature as a function of incident photon energies. For each valence band spectra, at a fixed binding energy position integrated intensity is computed with energy width of 0.6 eV which corresponds to minimum possible FWHM value of a photo emission line. Minimum FWHM value of a photo emission line is estimated from Au $ 4f $ spectra (data not shown here) and it is found to be $ \sim $0.6 eV. Obtain CIS plots for 5.0 eV and 4.2 eV valence band features mimic Ni and Mn $ 2p_{3/2} \rightarrow 3d $ photo absorption edges respectively, as presented in Figs. \ref{cisxasl3}(a, b). Therefore, CIS plots unambiguously confirm Ni and Mn $ d $ band characters of main resonance lines $ \beta $ and $ \zeta $, respectively. It is known that, ResPES studies contain information for both photo emission and photo absorption behaviors \cite{GvanderLaan1992I}. From CTM simulations, obtained $ L_{3} $ edge absorption spectra for Ni$ ^{2+/3+} $ and Mn$ ^{4+/3+} $ species are shown in Figs. \ref{cisxasl3}(a, b). Comparison of CIS plots with simulated XAS spectra again suggest mixed valency characters of both Ni, Mn transition metal ions, which is in agreement with XPS studies. 

Noticeably, for both Ni and Mn cases, valence band main resonance features are observed at higher binding energy positions, as highlighted in Figs. \ref{onoffrespes}(a, c). Other than these, presence of valence band features at lower binding energy side of main resonance characters are observed. Aforementioned CTM analysis on SNMO system confirms the presence of transition metal $ 3d $ - O $ 2p $ hybridized lower energy charge transfer screened $ d^{n+1}\underline{L} $ and higher energy unscreened $ d^{n} $ configurations mixing. Density functional theory (DFT) simulations on bulk ANMO family indicate significant overlapping between the O $ 2p $, Ni/Mn $ 3d $ states at valence band close to the Fermi level \cite{HDas2008, SKumar2010, HJZhao2014}. These studies transpire that in SNMO valence band, both Ni and Mn have low lying $ d^{n}\underline{L} $ final sates along with higher lying $ d^{n-1} $ final states. There are some previous studies based on 2\textit{p}$ \rightarrow $3\textit{d} ResPES experiments performed on prototype LNMO system. It was proposed that in LNMO, the valence band maximum has dominating Mn 3\textit{d} contribution \cite{MKitamura2009, JSKang2009}. However, DFT calculations on LNMO system indicate significant overlapping between the O $ 2p $, Ni $ 3d $ states at valence band maximum \cite{HDas2008}. Combining the results obtained from our present study, it is revealed that in SNMO system the top most of valence band comprises of O 2\textit{p} hybridized Ni 3\textit{d}$ ^{n} \underline{L} $ screened states. This is because of moderate covalent character, strong metal-ligand hybridization, and preferred charge transfer nature of Ni/Mn species in SNMO. Similar charge transfer behavior is reported for another prototype system Pr$ _{2} $NiMnO$ _{6} $ \cite{PBalasubramanian2018}. Observed broadening in $ \beta $ and $ \zeta $ main resonance features are due to octahedral crystal field splitting of transition metal $ d^{n-1} $ bands into low lying t$ _{2g} $ and higher lying e$ _{g} $ states. The signatures of t$ _{2g} $ and e$ _{g} $ characters are clearly distinguishable in feature $ \beta $ of Ni $ L_{3} $ ResPES (Fig.\ref{onoffrespes}(a)). On the other hand, from Mn $ L_{3} $ ResPES, the t$ _{2g} $ character can be identified in feature $ \zeta $ whereas, the e$ _{g} $ signature (at higher binding energy side) is not distinctly visible (Fig.\ref{onoffrespes}(b)). These differences in Ni and Mn t$ _{2g} $, e$ _{g} $ spectral weights are because of different electron population in different possible states of SNMO system as follows, Ni$ ^{2+} $: t$ _{2g}^{6} $ e$ _{g}^{2} $, Ni$ ^{3+} $(LS): t$ _{2g}^{6} $ e$ _{g}^{1} $, Ni$ ^{3+} $(HS): t$ _{2g}^{5} $ e$ _{g}^{2} $, Mn$ ^{4+} $: t$ _{2g}^{3} $ e$ _{g}^{0} $ and Mn$ ^{3+} $(HS): t$ _{2g}^{3} $ e$ _{g}^{1} $. Deconvolution of valence band spectra (discussed later) evidences the crystal field splitting of $ 3d $ bands into t$ _{2g} $ and e$ _{g} $ states. 

\begin{figure*}[t]
\centering
\includegraphics[angle=0,width=0.8\textwidth]{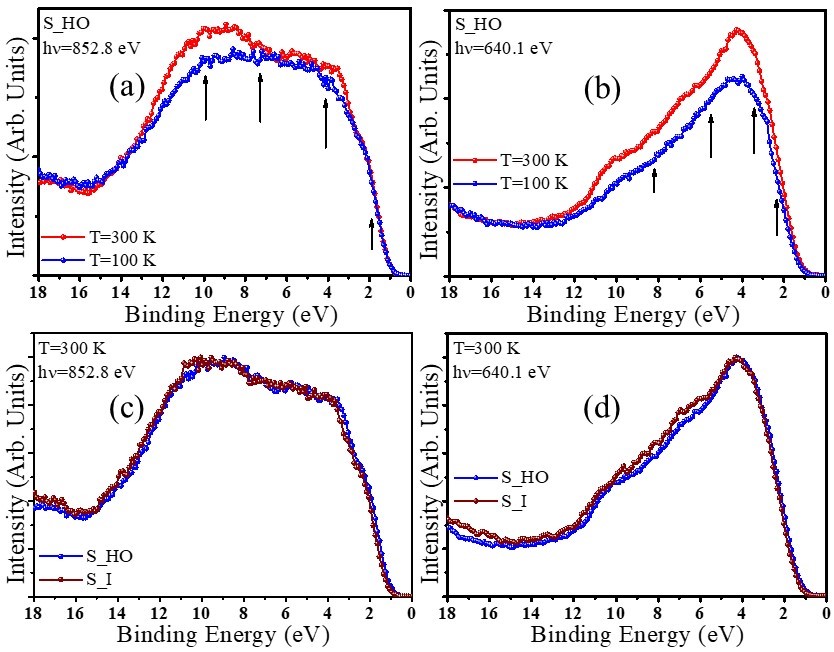}
\caption{On resonance valence band photo emission spectra measured for (a): S$ \_ $HO sample at T=300 K, 100 K across Ni, (b): S$ \_ $HO sample at T=300 K, 100 K across Mn, (c): S$ \_ $HO, S$ \_ $I samples at T=300 K across Ni and (d): S$ \_ $HO, S$ \_ $I samples at T=300 K across Mn $ 2p_{3/2} \rightarrow 3d $ absorption thresholds.}\label{ontasd}
\end{figure*}

With the photon energy variation most of the valence band features are stable in binding energy scale, as marked by upward arrows in Figs. S3-S4 in SM. This phenomena is often called as resonant Raman Auger scattering \cite{NMartensson1997}. Moreover, valence band features with constant kinetic energy $ \sim $844.7 eV for Ni and 637.5 eV for Mn cases, show dispersive nature in binding energy scale, as highlighted by open circles in Figs. S3-S4 in SM. These features are identified as Ni and Mn $ L_{3}VV $ regular Auger lines. The onset of regular Auger emission depends on corresponding element specie. In SNMO, Ni and Mn $ L_{3}VV $ Auger lines emerge in the vicinity of absorption thresholds at photon energy $ \sim $852.0 eV and 640.5 eV, respectively. The simultaneous occurrence of resonant Raman Auger and regular Auger process causes huge intensity enhancement of $ \gamma $ feature observed in Ni $ L_{3} $ ResPES. 

Temperature dependent Ni and Mn ResPES maps are displayed in Fig. S5 in SM. The variation in ResPES behavior for different measuring temperatures can also be distinguished in on-off resonance spectra, as illustrated by Fig. S6 in SM. Qualitatively similar trend is observed at T$ < $T$ _{C} $ (T=100 K, 125 K) ResPES studies as for the T$ > $T$ _{C} $ (T=300 K) case. Recorded on resonance valence band spectra at Ni and Mn $ L_{3} $ edges, are compared at different temperature values, as shown in Figs. \ref{ontasd}(a, b). With decreasing temperature valence band spectra becomes narrower and its spectral intensity is suppressed. In oxide materials, it is generally observed that O $ 2p $ valence band features enhance with increasing temperature \cite{DDSarma1996I}. Therefore, temperature dependent modifications of valence band structure suggest significant spectral weight from O $ 2p $ states. This observation is in accordance with DFT calculations on prototype ANMO system, which indicate mixing of O $ 2p $ - Ni/Mn $ 3d $ states in valence band \cite{HDas2008, SKumar2010, HJZhao2014}. Bandwidth narrowing of occupied level at low measurement temperatures is possibly because of decreased thermal broadening effects of photoemission lines. ResPES studies on SNMO samples having different level of ASD concentration show small broadening of valence band and resonance behaviors with increasing ASD fractions, as depicted in Figs. \ref{cisxasl3}(a, b) and Figs. \ref{ontasd}(c, d), respectively. However, with varying ASD densities, the positions of the spectral characters remain unaltered.

\begin{figure*}[t]
\centering
\includegraphics[angle=0,width=0.8\textwidth]{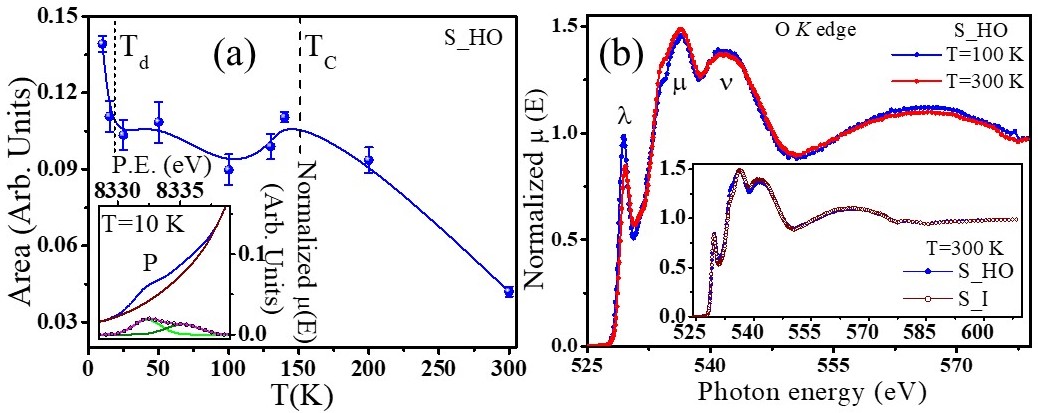}
\caption{(a): Thermal evolution of Ni $ K $ photo absorption pre-edge integrated intensity. Inset: Area under the curve of pre-edge structure measured at T=10 K. (b): Temperature dependent oxygen $ K $ edge photo absorption spectra for S$ \_ $HO thin film. Inset: Oxygen $ K $ edge absorption for different SNMO films with varying anti-site disorder extent recorded at T=300 K.}\label{nipreokedge}
\end{figure*}

Transition metal $ K $ absorption pre-edge feature involving excitation from 1$ s $ to empty 3$ d $ states, provides information about the local coordination environment and hybridization characters of the absorbing species. In general, these transitions are possibly electric quadrupole or dipole types \cite{FBridges2001}. For centrosymmetric absorbent site, $ 1s \rightarrow 3d $ dipole transitions are not allowed. However, distortion from centrosymmetry may cause hybridization between metal $ 3d-4p $ orbitals through ligand interactions and for those cases $ 1s \rightarrow 3d $ dipole transitions are weakly allowed \cite{FBridges2001}. On the other hand, the irreducible representations obtained from group theory calculation predicted that for octahedral point group, transition metal $ 3d-4p $ orbital mixing is forbidden \cite{TYamamoto2008}. As SNMO crystal structure is centrosymmetric (SG: \textit{P2$ _{1} $/n}) and  absorbent Ni is octahedrally coordinated with ligand O ions, observed pre-edge feature is attributed to $ 1s \rightarrow 3d $ electric quadrupole transition. Point to be noted here that transition metal $ K $ pre-edge features are also influenced by O 2$ p $ holes due to strong hybridization between metal 3$ d $ - O 2$ p $ states \cite{FBridges2001}. In a previous report on Mn \textit{K} edge XANES studies it was observed that with increasing Mn-O coordination distances the pre-edge structure changes, which can be correlated with the modification in Mn-O hybridization strength \cite{CGuglieri2011}. The temperature dependency of Ni \textit{K} pre-edge feature is displayed by Fig. S7 in SM. The area under the curve of pre-edge features are calculated using Gaussian peak shapes and arc-tangent background function, as presented by Inset of Fig. \ref{nipreokedge}(a). The thermal evolution of pre-edge integrated intensity shows distinct anomalies in the vicinity of magnetic ordering at T=T$ _{C} $ and T$ _{d} $ (Fig. \ref{nipreokedge}(a)). This trend suggests temperature driven modification in transition metal $ 3d $ - ligand $ 2p $ hybridization in SNMO system across the magnetic transitions.

Oxygen $ K $ near edge absorption spectra probes the unoccupied O $2p$ state hybridized with various states of constituent element present in the system \cite{JSuntivich2014, deGroot1989I}. O $K$ edge XANES spectra for SNMO thin films, measured at T=300 K are presented in Inset of Fig. \ref{nipreokedge}(b). The first feature $ \lambda $ at around 529.7 eV photon energy position is attributed to transition between O $ 1s $ and hybridized O $ 2p $ - Ni/Mn $ 3d $ states. Whereas, the second and third broad features $ \mu $ and $ \nu $ at $ \sim $536.2 eV and 541.8 eV respectively, correspond to excitation from O $ 1s $ to hybridized O $ 2p $ - Sm $ 5d $ and O $ 2p $ - Ni/Mn $ 4sp $ bands, respectively. The spectral shape and weight of feature $ \lambda $ depends on hybridization strength of O $ 2p $ - transition metal $ 3d $ levels and number of unoccupied states at those bands or in other words electron population of transition metal $ d $ states \cite{JSuntivich2014}. 

Due to hybridization, the octahedral crystal field from surrounding lattice O ligands, splits the transition metal \textit{d} bands into low lying t$ _{2g} $ and higher lying e$ _{g} $ states \cite{deGroot1989I}. In SNMO, as both Ni and Mn transition metals are octahedrally coordinated with O ions in similar geometry, the corresponding energy levels for t$ _{2g} $ and e$ _{g} $ states are closely spaced to each other, where each state have own finite energy broadening. This cause formation of combined band and visual identification of t$ _{2g} $ and e$ _{g} $ features are not observed. Deconvolution of O $ K $ edge spectra (discussed later) evidences the presence of crystal field splitted t$ _{2g} $ and e$ _{g} $ states of transition metal 3$ d $ bands. 

With decreasing temperature spectral weight of the feature $ \lambda $ in O $ K $ XANES increases, as presented in Fig.\ref{nipreokedge}(b). This confirms enhancement of O $ 2p $ - Ni/Mn $ 3d $ hybridization strength as measurement temperature is reduced below T$_C $. On the other hand, the possible presence O vacancy in lattice structure will modify the ligand field symmetry and electron hole ratio in transition metal \textit{d} band. Consequently, the spectral shape and intensity of feature $ \lambda $ is a sensitive probe for the lattice oxygen content of the system \cite{SMajumder2019, NBiskup2014}. We have observed that for different SNMO films with varying growth conditions, feature $ \lambda $ in O $ K $ XANES show similar spectral shape and weight (Inset of Fig. \ref{nipreokedge}(b)). Therefore, the extent of hybridization between O $ 2p $ - Ni/Mn $ 3d $ states and lattice oxygen content remain same in all SNMO films. This is because, the change in ASD densities does not affect the average crystal structure and within studied growth conditions there is no major O vacancy defect formation which can alter valency of constituent elements in SNMO system. 

\begin{figure*}[t]
\centering
\includegraphics[angle=0,width=1.0\textwidth]{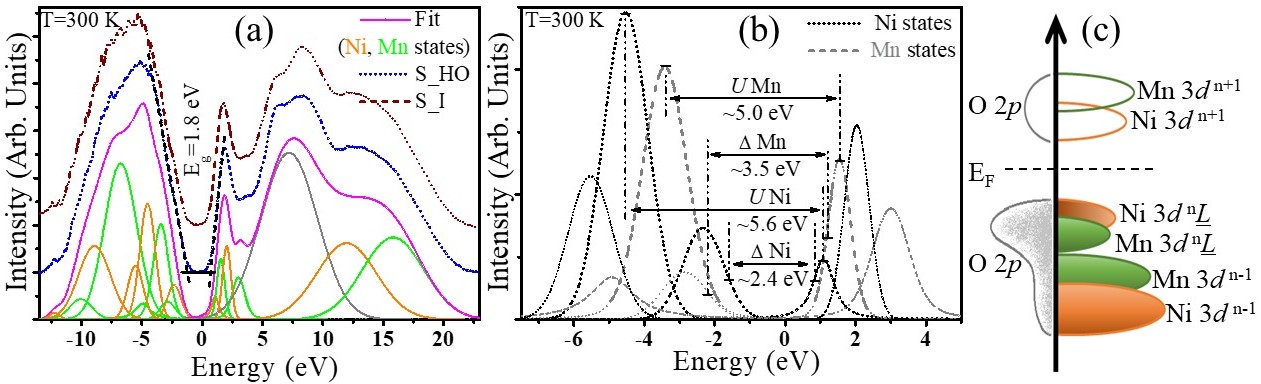}
\caption{(a): Experimental electronic band structure along with deconvoluted spectra across the Fermi level (E$ _{F} $) measured at T=300 K for SNMO films having different anti-site disorder concentrations. (b): Enlarged view of electronic states at valence and conduction bands in the vicinity of E$ _{F} $. (c): Schematic representation of overall band structure for SNMO system, showing Ni/Mn $ 3d $ and O $ 2p $ states near the E$ _{F} $ energy level.}\label{ebands}
\end{figure*}

Valence band spectra for $ h\nu $=634.6 eV and O $ K $ near edge absorption spectra are combined in common energy scale to construct the electronic band structure across the E$ _{F} $ \cite{TSaitoh1995II} at T=300 K, as illustrated in Fig. \ref{ebands}(a). For occupied band, photo emission spectra of excitation energy $ h\nu $=634.6 eV is chosen, because this incident photon energy satisfies the off resonance condition for both Ni/Mn species to avoid any abrupt change in spectral intensity by resonance effect. O $ K $ edge absorption spectra is used as unoccupied band, as it has dominating contribution from Ni/Mn transition metal $ 3d $ states. On the other hand, O $ 1s $ core hole effect on unoccupied band have lesser impact than transition metal case \cite{deGroot1989I}. The E$ _{F} $ alignment in valence band and conduction band is done by measuring binding energies for C 1\textit{s} ($ \sim $284.8 eV) and O 1\textit{s} ($ \sim $529.6 eV) features, respectively. Estimated band gap from tangent extrapolation at rising edges of valence band maxima and conduction band minima is found to be E$ _{g} $=1.8$ \pm $0.3 eV, for both S$ \_ $HO and S$ \_ $I samples. Accounting the error, this value has good resemblance with DFT predicted band gap E$ _{g} $=1.55 eV for SNMO bulk system \cite{HJZhao2014}. After removing inelastic background, deconvolution of SNMO band structure is carried out using minimum number of combined Gaussian-Lorentzian peak shapes, which adequately reproduce the spectral shape of observed spectra for S$ \_ $HO film. Qualitatively similar nature is found for S$ \_ $I sample (fitting not shown here). For background subtraction, Shirley and arc-tangent functions are used in valence band and conduction band regions, respectively. DFT studies on cation ordered bulk ANMO system suggest that the occupied band is composed of O $ 2p $, Ni and Mn $ 3d $ states, while unoccupied band is mainly consisted of Ni and Mn $ 3d $ states \cite{HDas2008, SKumar2010, HJZhao2014}. Aforementioned ResPES analysis reveal the presence of low energy $ d^{n}\underline{L} $ and high energy $ d^{n-1} $ states in valence band. Therefore, the valence band features close to E$ _{F} $ are attributed to O $ 2p $ hybridized Ni and Mn $ 3d $ states with low lying $ 3d^{n}\underline{L} $, higher lying $ 3d^{n-1} $ characters, respectively. Conduction band features near E$ _{F} $ are ascribe to Ni and Mn $ 3d^{n+1} $ states as discussed in O $ K $ XANES analysis. An enlarged view of contributing peaks across E$ _{F} $ is displayed in Fig. \ref{ebands}(b). These spectral features in the vicinity of E$ _{F} $ are ascribed to octahedral crystal field splitted low lying $ t_{2g} $ and higher lying $ e_{g} $ characters of Ni/Mn $ 3d $ states \cite{HDas2008, SKumar2010, HJZhao2014}. Spectroscopic value of $ \Delta _{p-d} $ is estimated from energy separation between band edges of $ 3d^{n}\underline{L} $ and $ 3d^{n+1} $ states. Band edge energy positions are defined as inflection point of corresponding peak feature. Obtained $ \Delta _{p-d} $ values for SNMO system are $ \Delta _{p-d} $(Ni)$ \sim $2.4 eV, $ \Delta _{p-d} $(Mn)$ \sim $3.5 eV. Spectroscopic value of $ U_{d-d} $ is estimated from the gap between lower and upper Hubbard bands, defined as peak centers of $ 3d^{n-1} $ and $ 3d^{n+1} $ features, respectively. For SNMO system $ U_{d-d} $ values are found to be $ U_{d-d} $(Ni)$ \sim $5.6 eV and $ U_{d-d} $(Mn)$ \sim $5.0 eV. Observed $ U_{d-d} > \Delta _{p-d} $ condition confirms charge transfer insulating nature of Ni/Mn cations in SNMO films. A schematic view depicting the overall electronic band structure for SNMO thin film system is illustrated in Fig. \ref{ebands}(c). Therefore, across E$ _{F} $ ligand to metal $ p-d $ type charge fluctuation is energetically favorable for both Ni, Mn transition metals. Temperature dependency of electronic band structure reveals (data not shown here) modification in spectral weights near E$ _{F} $ due to change in hybridization as already discussed in XANES analysis. Whereas, band structures are comparatively rigid with respect to variation in ASD concentration in SNMO system. 

Any physical phenomenon in transition metal oxides, for instance magnetic and electronic properties, are highly influenced by the competing interactions experienced by metal $d$ electrons. Such opposite forces mainly involve electron localization by Coulomb repulsion ($U_{d-d}$), electron delocalization by transition metal $d$ - ligand $p$ hybridization ($T_{pd\sigma}$) and charge transfer ($\Delta _{p-d}$). Based on these interactions, transition metal compounds are classified as $d$ or $p$ type metals and Mott Hubbard ($U_{d-d} < \Delta _{p-d}$) or Charge transfer ($U_{d-d} > \Delta _{p-d}$) insulators \cite{JZaanen1985}. Cation ordered ANMO double perovskite structure can be approximated as combination of ANiO$ _{3} $ and AMnO$ _{3} $ single perovskite stacking in alternate fashion. ANiO$_3$ nickelates and AMnO$_3$ manganites are categorized respectively as negative Charge transfer insulator with $ p-p $ ($ d^{n+1}\underline{L} + d^{n+1}\underline{L} \rightarrow d^{n+1}\underline{L}^{2} + d^{n+1} $) charge fluctuation and Mott Hubbard insulator with $ d-d $ ($ d^{n} + d^{n} \rightarrow d^{n+1} + d^{n-1} $) charge fluctuation \cite{SMiddey2016, AChainani1993}. In earlier studies, SmNiO$ _{3} $ is observed to show AFM and M-I transitions at T$ _{N} \simeq $ 225 K and T$ _{M-I}(Ni) \simeq $ 400 K, respectively \cite{MLMedarde1997}. On the other hand, SmMnO$ _{3} $ is reported to exhibit AFM-I state with magnetic transition temperature T$ _{N}(Mn) \simeq $ 59 K \cite{TKimura2003}. However, for SNMO system, we have observed that magnetic and electronic properties are very different from either SmNiO$ _{3} $ or SmMnO$ _{3} $ cases. SNMO thin film reveals FM-I state with $ p-d $ ($ d^{n} + d^{n} \rightarrow d^{n} + d^{n+1}\underline{L} $) low energy charge fluctuation. Moreover, as here, the magnetic properties are governed by intercompeting interactions from cation ordered (Ni-O-Mn) and disordered (Ni-O-Ni or Mn-O-Mn) structures, a proper control over Ni/Mn ASD density can effectively tune the magnetic states. Whereas, the electronic properties depend on overall concentration of Ni$ ^{2+/3+} $ and Mn$ ^{4+/3+} $ states which are found to be insensitive with respect to ASD variation. Therefore, present work establishes SNMO double perovskite epitaxial thin film as a FM-I system where FM state can be tuned by engineering ASD fraction in the crystal structure, while I state remains unaffected.

\section{CONCLUSION}
In summary, we have probed the alteration in electronic structure across the magnetic transition of epitaxial Sm$ _{2} $NiMnO$ _{6} $ double perovskite thin films having different level of cation ordering. SNMO system undergoes paramagnetic to long range ferromagnetic phase transition at T=T$ _{C} $. Anti-site disorder in the form of Ni-O-Ni or Mn-O-Mn bonds leads to short scale antiferromagnetic couplings in background of ferromagnetic interactions from Ni-O-Mn cation ordered structures. The inter-competing interactions from these coexisting ferromagnetic-antiferromagnetic phases govern the temperature driven magnetic phase diagram in SNMO. Increase in anti-site disorder concentration eventually results in shift of transition temperatures toward lower values, drastic reduction of saturation moment, decrease of remanence magnetization, drop of anisotropy energy and deviation of anisotropy nature from uniaxial to bi-axial type. Both Ni and Mn transition metals exhibit mixed valency due to $ Ni^{2+}+Mn^{4+} \longrightarrow Ni^{3+}+Mn^{3+} $ kind of charge disproportionation between cation sites in SNMO irrespective of different anti-site disorder density. Configuration interaction cluster model analysis of core level spectra suggests mixing of unscreened $ d^{n} $ and charge transfer screened $ d^{n+1} \underline{L}^{1} $ characters in SNMO system. Resonant photo emission measurements across Ni, Mn $ 2p \rightarrow 3d $ photo excitation regimes unambiguously confirm the presence of Ni/Mn low lying $ 3d^{n}\underline{L} $ ($\underline{L} $: O $ 2p $ hole), higher lying $ 3d^{n-1} $ final states with considerable O $ 2p $ weightage in valence band. Strong hybridization between Ni/Mn $ 3d $ - O $ 2p $ states owing to significant orbital overlapping, give rise aforementioned charge transfer characters in SNMO system. Electronic band structure analysis reveals that the occupied and unoccupied levels in the vicinity of $ E_{F} $ consist of mixed O $ 2p $, Ni and Mn $ 3d $ states. Spectroscopic estimation of Coulomb repulsion ($ U_{d-d} $), charge transfer ($ \Delta_{p-d} $) and band gap ($ E_{g} $) energy values confirm charge transfer insulating band gap with $ p-d $ type low energy charge fluctuation in SNMO thin films. Below magnetic ordering temperatures (T$ < $T$ _{C} $), enhancement of Ni/Mn $ 3d $ - O $ 2p $ hybridization strength is evidenced. Anti-site disorder results in slight broadening in Ni, Mn $ 3d $ spectral features, while the spectral positions are found to be rigid with respect to variation in cation arrangements. Present work establishes the importance of metal - ligand hybridization and cation distribution defect in understanding intriguing physical aspect of the ferromagnetic insulator system. We hope, this study will have huge relevance in FM-I system as quantum electronic materials. 
\\
\section*{ACKNOWLEDGMENTS}
Authors gratefully acknowledge Dr. Federica Bondino (IOM CNR, Italy) for fruitful discussions regarding the spectroscopic experiments performed at Elettra. Thanks to Elettra-Sincrotrone, Italy and Indus Synchrotron RRCAT, India for giving access to experimental facilities. Authors acknowledge the Department of Science and Technology, Government of India; Indian Institute of Science, Italian Government and Elettra for providing financial support through Indo-Italian Program of Cooperation (No. INT/ITALY/P-22/2016 (SP)) to perform experiments at Elettra-Sincrotrone. S.M. thanks Mr. Avinash Wadikar (CSR, India), Mr. Sharad Karwal (CSR), and Mr. Rakesh K. Sah (CSR) for their technical help in measurements at RRCAT. I.P. and S.N. gratefully acknowledge financial support from EUROFEL project (RoadMap Esfri).

\section*{REFERENCES}
\bibliography{}

\end{document}